\documentclass[a4paper,11pt]{article}
\usepackage[skins]{tcolorbox}
\usepackage{mathtools,slashed}
\usepackage[T1]{fontenc}
\usepackage{slashed,verbatim,subfigure}
\usepackage[numbers,sort&compress]{natbib}
\usepackage{amsmath}
\usepackage{mathrsfs}
\usepackage{amsbsy}
\usepackage{amssymb}
\usepackage{appendix}
\usepackage{graphicx}
\usepackage[final]{pdfpages}
\usepackage{dcolumn}
\usepackage{bm}
\usepackage[colorlinks=true,linktocpage=true,
linkcolor=blue,citecolor=blue]{hyperref}
\usepackage[a4paper]{geometry}
\catcode`@11
\def\seceqaa{\@addtoreset{equation}{section}
	\def\theequation{A\arabic{equation}}}
\def\seceqbb{\@addtoreset{equation}{section}
	\def\theequation{B\arabic{equation}}}
\def\seceqcc{\@addtoreset{equation}{section}
	\def\theequation{C\arabic{equation}}}
\def\seceqdd{\@addtoreset{equation}{section}
	\def\theequation{D\arabic{equation}}}
\def\seceqee{\@addtoreset{equation}{section}
	\def\theequation{E\arabic{equation}}}
\catcode`@11
\newcommand{\be}{\begin{eqnarray}}
\newcommand{\ee}{\end{eqnarray}}
\begin{document}
\large
\title{Deconfinement Temperature of Rotating QGP at Intermediate Coupling from ${\cal M}$-Theory}
\author{Gopal Yadav\footnote{email- gyadav@ph.iitr.ac.in}\vspace{0.1in}\\
Department of Physics,
Indian Institute of Technology Roorkee, Roorkee 247667, India}
\date{}
\maketitle
\begin{abstract}
With the aim of studying rotating quark-gluon plasma (QGP), holographically, from a top-down approach, the study of the effect of rotation on the deconfinement temperature of thermal QCD-like theories at intermediate coupling from ${\cal M}$-theory was missing in the literature. This paper fills this gap. The gravity dual includes a rotating cylindrical black hole. In the presence of rotation, from a semi-classical computation, we found that the deconfinement temperature is inversely proportional to the Lorentz factor, which suggests that the deconfinement temperature decreases with the increase of rotation. Further, we found that in the small angular velocity limit, results from higher derivative correction at ${\cal O}(R^4)$ do not change and are the same as in \cite{McTEQ}. The ``UV-IR mixing'', ``Flavor Memory'' effect, and ``non-renormalization of $T_c$'' in the ${\cal M}$-theory dual are similar to the ones observed in \cite{McTEQ}.
\end{abstract}

\section{Introduction}
\label{introduction}
AdS/CFT duality relates strongly coupled gauge theories and weakly coupled gravitational theories. In \cite{AdS/CFT}, the author realized that strongly coupled ${\cal N}=4$ SYM theory and type IIB supergravity background on $AdS_5 \times S^5$ are dual to each other. Using gauge-gravity duality, we can study many interesting phenomena in strongly coupled gauge theories via a mapping to the gravitational dual. In the gravitational dual, calculations are easily doable, making gauge-gravity duality more helpful in studying the strongly coupled gauge theories. If we incorporate higher derivative terms in the holographic dual of strongly coupled gauge theories, we can explore the intermediate coupling regime of the gauge theories. The effect of higher derivative corrections in gravity dual of ${\cal N}=4$ SYM theory at finite temperature was studied in \cite{HD-SYM}.
 
\par  Since QCD is a strongly coupled gauge theory in the low-energy regime and weakly coupled in the high-energy regime, we can hope to study QCD via an appropriate generalization of the AdS/CFT duality. We can study any gauge theory via gauge-gravity duality from the top-down and bottom-up approaches. There are models constructed based on a bottom-up approach e.g., hard-wall model \cite{HW1} and soft-wall model \cite{SWM}. Gravity duals in the models, as mentioned above, involve an AdS background. We will study thermal QCD from a top-down approach. There are two known models from the top-down perspective that we are aware of, the Sakai-Sugimoto model \cite{SS}, which caters to the IR only, and the model constructed in \cite{metrics,MQGP,NPB} which is UV complete. In \cite{MQGP}, authors were able to explore thermal QCD-like theories at the finite coupling. In \cite{HD-Roorkee}, authors incorporated ${\cal O}(R^4)$ terms in the ${\cal M}$-theory uplift constructed in \cite{MQGP} such that one can get analytical control on the thermal QCD-like theories at the intermediate coupling.
\par
We can calculate the deconfinement temperature of thermal QCD using gauge-gravity duality from a semi-classical computation \cite{Witten-Tc} by equating on-shell action of the black hole and thermal backgrounds at UV cut-off. Authors in \cite{Tc-HW_SW} calculated the deconfinement temperature of holographic QCD using the hard-wall model \cite{HW1} and soft-wall model \cite{SWM}. We calculated the deconfinement temperature of thermal QCD-like theories in the absence of rotation at intermediate coupling in \cite{McTEQ}, and in the same process we demonstrated a novel ``UV-IR'' mixing in our setup. In any theory, we characterize the low-energy regime as the IR regime and the high-energy regime as the UV regime. Under renormalization group flow, we go from the ultraviolet regime of the theory to the infrared regime. In string theory, physics at the UV scale is related to physics at the IR scale, which is defined as ``UV-IR'' mixing and it was discussed in the context of AdS/CFT correspondence in \cite{UV-IR-AdS-CFT}.
\par
Quark Gluon Plasma (QGP) produced in noncentral relativistic heavy ion collisions (RHIC), is a highly vortical fluid. Experimental results of RHIC collaboration \cite{RHIC-omega} say that this vortical fluid has an angular momentum of the order of $10^3 \hbar$ and has an angular velocity of the order of $\omega \sim 6 MeV$. In \cite{omega-simulation}, authors predicted $\omega \sim 20-40 MeV$ from hydrodynamics simulations of heavy ion collisions. Based on the aforementioned results, it will be very interesting to study the effect of rotation on the deconfinement temperature of thermal QCD-like theories using gauge-gravity duality from a top-down approach. Authors in \cite{R-Tc}, studied the effect of rotation on the deconfinement temperature of QCD from hard wall and soft wall models, and they found that as the angular velocity of the plasma is increasing deconfinement temperature of QCD decreases. Similarly, there are other works from the bottom-up approach \cite{G-D-Rotation,Kerr-AdS-1,Kerr-AdS-2} where similar result appeared.

 We have organized the paper in the following way. In section \ref{review}, we discuss the top-down holographic dual of rotating QGP at the intermediate coupling via two subsections \ref{M-Theory-Uplift} and \ref{Rotating-QGP}. We compute the deconfinement temperature of thermal QCD at intermediate coupling in \ref{O-beta0-Tc} and discuss ``UV-IR'' mixing, ``Flavor Memory'' effect, and ``non-renormalization of $T_c$'' in \ref{UV-IR-mixing}. We summarise our results in section \ref{summary}. In appendix \ref{metric}, we have listed the metric components of the ${\cal M}$-theory involving a rotating cylindrical black hole background and in appendix \ref{beta}, we obtain ${\cal O}(\beta)$ contribution to the on-shell action densities for rotating cylindrical black hole and thermal backgrounds.

\section{Top-Down Holographic Dual of Rotating QGP at Intermediate Coupling}\label{review}
As for as we know, the string dual of thermal QCD-like theories at finite temperature in the large-$N$ limit was constructed in \cite{metrics}, which exhibits ultraviolet(UV) completion. Brane setup of \cite{metrics} involve $N$ color $D3$ branes, $M$  $D5$ and $\overline{D5}$ branes, $N_f$ flavor $D7$ and $\overline{D7}$ branes (embedded via Ouyang embedding \cite{Ouyang}) and the gravity dual is a resolved warped deformed conifold. In this model, we have infinite 't Hooft coupling on the gauge theory side, which is thermal QCD in our case. To study the thermal QCD-like theories at finite coupling, authors in \cite{MQGP} constructed type IIA S(trominger) Y(au) Z(aslow) mirror (for which we require a (delocalized) special Lagrangian (sLag) $T^3$, which is the $T^2$-invariant sLag of \cite{M.Ionel and M.Min-OO (2008)} with a large base ${\cal B}(r,\theta_1,\theta_2)$ (of a $T^3(x,y,z)$-fibration over ${\cal B}(r,\theta_1,\theta_2)$) \cite{NPB,EPJC-2}) of the type IIB setup and then uplifted it to the ${\cal M}$-theory. In the type IIA mirror, all the branes of the type IIB setup are converted into color and flavor $D6$-branes, and these branes will be replaced by geometry and fluxes (e.g., $D6$-branes converted into KK-monopoles) upon uplifting the type IIA mirror to ${\cal M}$-theory, and we have the gravity dual as ``${\cal M}$- theory on a $G_2$-structure manifold''. Under RG group flow from UV-to-IR and after the application of repeated Seiberg-like dualities, we are left with confining $SU(M)$ gauge theory in the IR, and $M$ can be tuned to $N_c=3$ in the MQGP limit (in general, $N_c = N_{\rm effective}+M_{\rm effective}$, if we represent  ``$a$'' as $[a_{UV},a_{IR}]$, then $N_{\rm eff} \in[N,0]$ and $M_{\rm eff} \in [0,M]$ implies $N_c \in [N, M]$ and hence the number of colors in the IR, $N_c=M$ \cite{McTEQ}) \cite{Misra+Gale}. Further, there was no analytical control on the thermal QCD-like theories at the intermediate coupling regime. To do so, authors in \cite{HD-Roorkee} incorporated ${\cal O}(R^4)$ terms on the gravity dual side. As a result, in terms of gauge-gravity duality, it has become possible to control theories such as thermal QCD at intermediate coupling analytically. We reviewed these works in our previous paper \cite{MChPT,McTEQ}. In this paper, we need the metric of ${\cal M}$-theory dual of thermal QCD-like theories and the same is explained in subsection \ref{M-Theory-Uplift} and listed in appendix \ref{metric}. For more details of the model, see section {\bf 2} of the papers \cite{MChPT, McTEQ}. 

\subsection{ ${\cal M}$-Theory Uplift at Intermediate Coupling}
\label{M-Theory-Uplift}
In this part, we briefly review how authors in \cite{HD-Roorkee}  obtained the ${\cal O}(R^4)$ correction to the ${\cal M}$-theory metric \cite{MQGP,NPB}. Eleven-dimensional supergravity action with higher derivative terms is given below
{\footnotesize
\begin{eqnarray}
\label{D=11_O(l_p^6)}
& & \hskip -0.5inS = \frac{1}{2 \kappa_{11}^2 } \Biggl[\int_{M_{11}} d^{11}x \sqrt{-G^{\cal M}}  {\cal R} - \frac{1}{2}\int_{M_{11}}  d^{11}x\sqrt{-G^{\cal M}}G_4^2 -
\frac{1}{6}\int_{{M_{11}}} C_3\wedge G_4\wedge G_4 \nonumber\\
& & \hskip -0.5in +2 \int_{\partial {M_{11}}} d^{10}x \sqrt{-h^{\cal M}} K \Biggr]+\beta  \Biggl( \int_{M_{11}}  d^{11}x \sqrt{-G^{\cal M}}\left( J_0 - \frac{1}{2}E_8\right) + 3^2 2^{13} \int_{M_{11}} C_3\wedge X_8\Biggr),
\end{eqnarray}
}
where $\beta \equiv \frac{(4\pi\kappa_{11}^2)^{2/3}}{(2\pi)^4 3^2 2^{13}} =\frac{(2 \pi)^2 l_p^6}{3^2 2^{13}}$, because, $\kappa_{11}^2=\frac{(2\pi)^8 l_p^9}{2}$ ($l_p$ is the eleven dimensional Planckian length) \cite{Becker-sisters-O(R^4),HD-Roorkee}, $C_3$ is the ${\cal M}$-theory three-form potential, $G_4=dC_3$, and
\begin{eqnarray}
\label{J0+E8-definitions}
& & \hskip -0.8inJ_0  =3\cdot 2^8 (R^{HMNK}R_{PMNQ}{R_H}^{RSP}{R^Q}_{RSK}+
{1\over 2} R^{HKMN}R_{PQMN}{R_H}^{RSP}{R^Q}_{RSK}),\nonumber\\
& & \hskip -0.8inE_8  ={ 1\over 3!} \epsilon^{ABCM_1 N_1 \dots M_4 N_4}
\epsilon_{ABCM_1' N_1' \dots M_4' N_4' }{R^{M_1'N_1'}}_{M_1 N_1} \dots
{R^{M_4' N_4'}}_{M_4 N_4}.
\end{eqnarray}
The equations of motion of the ${\cal M}$-theory metric and three-form potential are
{\footnotesize
\begin{eqnarray}
\label{eoms}
& & \hskip -0.1in R_{MN} - \frac{1}{2}G^{\cal M}_{MN}{\cal R} - \frac{1}{12}\left(G_{MPQR}G_N^{\ PQR} - \frac{G^{\cal M}_{MN}}{8}G_{PQRS}G^{PQRS} \right)  = - \beta\left[\frac{G^{\cal M}_{MN}}{2}\left( J_0 - \frac{1}{2}E_8\right) + \frac{\delta}{\delta G_{\cal M}^{MN}}\left( J_0 - \frac{1}{2}E_8\right)\right],\nonumber\\
& & d*G = \frac{1}{2} G\wedge G +3^22^{13} \left(2\pi\right)^{4}\beta X_8,
\end{eqnarray}
}
where $R_{MNPQ}, R_{MN}, {\cal R}$  in  (\ref{D=11_O(l_p^6)})/(\ref{eoms}) are Riemann curvature tensor, Ricci tensor and the Ricci scalar in the eleven dimensions. Authors in \cite{HD-Roorkee} made the following ansatz for ${\cal M}$-theory metric and three-form potential
\begin{eqnarray}
\label{ansaetze}
& & \hskip -0.8inG^{\cal M}_{MN} = G_{MN}^{(0)} +\beta G_{MN}^{(1)}; \ C_{MNP} = C^{(0)}_{MNP} + \beta C_{MNP}^{(1)}.
\end{eqnarray}
It was explained in \cite{HD-Roorkee} that $C^{(1)}_{MNP}=0$ up to ${\cal O}(\beta)$. Therefore correction will be only to the ${\cal M}$-theory metric and defined as
\begin{eqnarray}
\label{fMN-definitions}
\delta G^{\cal M}_{MN} =\beta G^{(1)}_{MN} = G^{\rm MQGP}_{MN} f_{MN}(r),
\end{eqnarray}
where there is no summation in the above equation, by using equations (\ref{eoms}) and (\ref{fMN-definitions}), authors obtained the equations of motion of the $f_{MN}(r)$ for the black hole background and their solutions in \cite{HD-Roorkee}. Similarly, we obtained ${\cal O}(R^4)$ correction for the ${\cal M}$-theory thermal background in \cite{MChPT} where we assumed the blackening function for the thermal background equal to ``one''. It was explained in \cite{HD-Roorkee} and reviewed in \cite{McTEQ} that we can not go beyond ${\cal O}(\beta)$.\\ \\
{\bf On-Shell Action:} We worked out the on-shell action of (\ref{D=11_O(l_p^6)}) in \cite{McTEQ} and is given below (in $\kappa_{11}^2=1$ unit)
{\footnotesize
\begin{equation}
\label{on-shell-D=11-action-up-to-beta}
 S_{D=11}^{\rm on-shell} = -\frac{1}{2}\Biggl[-2 S_{\rm EH}^{(0)} + 2 S_{\rm GHY}^{(0)} 
+ \beta \left(\frac{20}{11}S_{\rm EH} - 2 \int_{M_{11}}\sqrt{-G_{\cal {M}}^{(1)}}R^{(0)}
+ 2 S_{\rm GHY} - \frac{2}{11}\int_{M_{11}}\sqrt{-G^{\cal {M}}}G^{MN}\frac{\delta J_0}{\delta G^{MN}}\right)\Biggr].
\end{equation}
}
To calculate the various terms appearing in equation (\ref{on-shell-D=11-action-up-to-beta}), we need to write all the quantities as ${\cal A}={\cal A}^{\rm MQGP}\left(1+ \beta f_{MN}\right)$, where ${\cal A}^{\rm MQGP}$ is the quantity ${\cal A}$ in the ``MQGP'' limit (MQGP limit is defined as: $g_s\sim\frac{1}{{\cal O}(1)}, M, N_f \equiv {\cal O}(1),\ g_sN_f<1,\ N\gg1,\ \frac{g_s M^2}{N}\ll1,$
which requires finite string coupling; therefore, we can address the MQGP limit from ${\cal M}$-theory \cite{MQGP,NPB}), i.e., without the inclusion of ${\cal O}(R^4)$ correction and ${\cal A}^{\rm MQGP}f_{MN}$, is the correction to the aforementioned quantity coming from higher derivative term. In particular,
\begin{eqnarray}
\label{metric-HD}
& & \hskip -0.5in
G^{\cal {M}}_{MN}=G^{\rm MQGP}_{MN}\left(1+ \beta f_{MN}\right);  R=R^{(0)}+\beta R^{(1)}; h^{\cal {M}}_{MN}=h^{\rm MQGP}_{MN}\left(1+ \beta f_{MN}\right); K=K^{(0)}+\beta K^{(1)},
\end{eqnarray}
implying $ S_{\rm EH}^{\cal M}=S_{\rm EH}^{(0)}+\beta S_{\rm EH}^{(1)}$ and $S_{\rm GHY}^{\cal M}=S_{\rm GHY}^{(0)}+\beta S_{\rm GHY}^{(1)}$. The metric of the ${\cal M}$-theory dual for the deconfined phase of thermal QCD-like theories ($T>T_c$) is \cite{McTEQ}:
\begin{eqnarray}
\label{TypeIIA-from-M-theory-Witten-prescription-T>Tc}
\hskip -0.2in ds_{11}^2|_{BH} & = & e^{-\frac{2\phi^{\rm IIA}}{3}}\Biggl[\frac{1}{\sqrt{h(r,\theta_{1,2})}}\left(-g(r) dt^2 + \left(dx^1\right)^2 +  \left(dx^2\right)^2 +\left(dx^3\right)^2 \right)
\nonumber\\
& & \hskip -0.2in+ \sqrt{h(r,\theta_{1,2})}\left(\frac{dr^2}{g(r)} + ds^2_{\rm IIA}(r,\theta_{1,2},\phi_{1,2},\psi)\right)
\Biggr] + e^{\frac{4\phi^{\rm IIA}}{3}}\left(dx^{11} + A_{\rm IIA}^{F_1^{\rm IIB} + F_3^{\rm IIB} + F_5^{\rm IIB}}\right)^2,
\end{eqnarray}
where $g(r)=1-\frac{r_h^4}{r^4}$, $\phi^{\rm IIA}$ is the type IIA dilaton profile, $h(r,\theta_{1,2})$ is the warp factor in ten dimensions \cite{metrics,MQGP} and $A_{\rm IIA}^{F_{i=1,3,5}^{\rm IIB}}$ are RR 1-forms in type IIA string theory constructed from type IIB fluxes ($F_{i=1,3,5}^{\rm IIB}$) \cite{MQGP}. Geometry of (\ref{TypeIIA-from-M-theory-Witten-prescription-T>Tc}) is $S^1(t) \times_w \mathbb{R}^{3}(x^{1,2,3}) \times_w \mathbb{R}_{\geq 0}(r) \times_w {\cal M}_6(\theta_{1,2},\phi_{1,2},\psi,x^{10})$, where ``$\times_w$'' denotes warped product. Periodicity of thermal circle is, $\beta_{\rm BH} =\frac{1}{T_H} \sim \frac{L^2}{r_h} \sim \frac{\sqrt{N}}{e^{-\kappa_{r_h}N^{1/3}}}$ because $L^4=4\pi g_s N$ and $r_h \sim e^{-\kappa_{r_h}N^{1/3}}$ \cite{Bulk-viscosity}. Hence, in the large-$N$ limit, $\beta_{\rm BH}$ will be very large and one can treat thermal circle as non-compact direction. Further, in the UV, $h(r) \sim \frac{L^4}{r^4}$, and $\phi^{\rm IIA}$ is constant, therefore (\ref{TypeIIA-from-M-theory-Witten-prescription-T>Tc}) has the following form in the UV
\begin{eqnarray}
\label{conformal-metric-BH}
& &
{ds^2_{11}|}_{\rm BH}^{\rm UV} \approx \frac{r^2}{L^2}\Biggl(-g(r) dt^2 + \left(dx^1\right)^2 +  \left(dx^2\right)^2 +\left(dx^3\right)^2 +\frac{L^4}{r^4}\left(\frac{dr^2}{g(r)}\right)\Biggr)\nonumber\\
& & + \Biggl(\frac{L^4}{r^4} ds^2_{\rm IIA}(r,\theta_{1,2},\phi_{1,2},\psi)+e^{\frac{4\phi^{\rm IIA}}{3}}\left(dx^{11} + A_{\rm IIA}^{F_1^{\rm IIB} + F_3^{\rm IIB} + F_5^{\rm IIB}}\right)^2 \Biggr).
\end{eqnarray}
Therefore, in the large-$N$ limit, conformal boundary of the metric (\ref{TypeIIA-from-M-theory-Witten-prescription-T>Tc}), i.,e., when $r \rightarrow {\mathcal {R}}_{\rm UV}$ is $AdS_5(\mathbb{R}^{1,3}(t,x^{1,2,3}), \mathbb{R}_{\geq0}(r)) \times_w {\cal M}_6(\theta_{1,2},\phi_{1,2},\psi,x^{10})$

\subsection{Introducing Rotation in Thermal QCD-Like Theories}
\label{Rotating-QGP}
 One can study the effect of rotation in thermal QCD-like theories by introducing rotating cylindrical black hole and thermal backgrounds on the gravity dual side when $T>T_c$ and $T<T_c$ on the gauge theory side. To obtain the rotating cylindrical black hole background, we need to make $x^3$ periodic by replacing it with $l \phi$, where $l$ is the length of the cylinder and $0\leq \phi \leq 2 \pi$. One can obtain the holographic dual of rotating quark-gluon plasma by performing the following Lorentz transformation around the cylinder of length $l$ \cite{Lorentz-boost-1,Lorentz-boost-2}.
\begin{eqnarray}
\label{Lorentz-boost}
& & t \rightarrow \frac{1}{\sqrt{1-l^2 \omega^2}}\left(t+l^2 \omega \phi \right); \ \phi \rightarrow \frac{1}{\sqrt{1-l^2 \omega^2}} \left(\phi + \omega t\right),
\end{eqnarray}
where $\omega$ is the angular velocity of the rotating cylindrical black hole in the gravity dual, which is associated with the rotation of quark-gluon plasma via gauge-gravity duality. Lorentz transformation (\ref{Lorentz-boost}) is valid only when $\omega l <1$. Under Lorentz transformation defined in equation (\ref{Lorentz-boost}), equation(\ref{TypeIIA-from-M-theory-Witten-prescription-T>Tc}) becomes:
{\footnotesize 
\begin{eqnarray}
\label{TypeIIA-from-M-theory-Witten-prescription-T>Tc-Rotation-Canonical}
\hskip -6in ds_{11}^2|_{BH} & = & e^{-\frac{2\phi^{\rm IIA}}{3}}\Biggl[\frac{1}{\sqrt{h(r,\theta_{1,2})}}\Biggl(-{\cal Y}_1(r) dt^2 +{\cal Y}_2(r)\left(d \phi+{\cal Y}_3(r) dt\right)^2 
 + \left(dx^1\right)^2 +  \left(dx^2\right)^2 \Biggr)
\nonumber\\
& & \hskip -0.1in+ \sqrt{h(r,\theta_{1,2})}\left(\frac{dr^2}{g(r)} + ds^2_{\rm IIA}(r,\theta_{1,2},\phi_{1,2},\psi)\right)
\Biggr] + e^{\frac{4\phi^{\rm IIA}}{3}}\left(dx^{11} + A_{\rm IIA}^{F_1^{\rm IIB} + F_3^{\rm IIB} + F_5^{\rm IIB}}\right)^2,
\end{eqnarray}
}
where ${\cal Y}_1(r)=\frac{g(r)\left(1-l^2 \omega^2\right)}{\left(1-g(r)l^2\omega^2\right)}; \
{\cal Y}_2(r)=\frac{l^2 \left(1- g(r)l^2 \omega^2\right)}{\left(1-l^2\omega^2\right)}; \
{\cal Y}_3(r)=\frac{\omega \left(1- g(r)\right)}{\left(1-g(r) l^2\omega^2\right)}$. The conformal boundary of the metric (\ref{TypeIIA-from-M-theory-Witten-prescription-T>Tc-Rotation-Canonical}) in the large-$N$ limit will be $AdS_4(\mathbb{R}^{1,2}(t,x^{1,2}),\mathbb{R}_{\geq0}(r)) \times_w {\cal M}_7(\phi,\theta_{1,2},\phi_{1,2},\psi,x^{10})$ ($\times_w$ implies warped product), but we have off-diagonal components in $(t,\phi)$ subspace (\ref{TypeIIA-from-M-theory-Witten-prescription-T>Tc-Rotation-Canonical}) in comparison to the static black hole metric (\ref{TypeIIA-from-M-theory-Witten-prescription-T>Tc}). One can diagonalize the metric in $(t,\phi)$ subspace in a new basis $(T,\Phi)$, which we did for ${\cal O}(\beta)$ correction in appendix \ref{beta} and then the conformal boundary of the rotating cylindrical black hole metric look like (\ref{conformal-metric-BH}) but now $AdS_4(\mathbb{R}^{1,2}(t,x^{1,2}), \mathbb{R}_{\geq0}(r)) \times_w {\cal M}_7(\phi,\theta_{1,2},\phi_{1,2},\psi,x^{10})$ instead of $AdS_5(\mathbb{R}^{1,3}(t,x^{1,2,3}), \mathbb{R}_{\geq0}(r)) \times_w {\cal M}_6(\theta_{1,2},\phi_{1,2},\psi,x^{10})$ in (\ref{conformal-metric-BH}) because of compact $\phi$ coordinate. There are no branes in the ${\cal M}$-theory uplift constructed in \cite{MQGP,HD-Roorkee} because when we uplift the type IIA mirror of type IIB setup \cite{metrics} then branes of type IIA string dual (color and flavor $D6$-branes) get converted into geometry and fluxes in the ${\cal M}$-theory uplift. Finally, we left with  ${\cal M}$-theory on a $G_2$-structure manifold. But we can make an analogy of the metric (\ref{TypeIIA-from-M-theory-Witten-prescription-T>Tc-Rotation-Canonical}) with black $M2$-brane due to its structure (because of two spatial and one time coordinate in (\ref{TypeIIA-from-M-theory-Witten-prescription-T>Tc-Rotation-Canonical})). We would look to make it clear that this is just an analogy because ${\cal M}$-theory uplift is a ``no-braner'' uplift.

Now using equation (\ref{TypeIIA-from-M-theory-Witten-prescription-T>Tc-Rotation-Canonical}), we can calculate Hawking temperature of the rotating cylindrical black hole using the formula given in \cite{G-D-Rotation}
\begin{eqnarray}
\label{Hawking-temp-defn}
T_H(\gamma)=\Biggl|\frac{\kappa}{2 \pi}\Biggr|=\Biggl|\frac{\lim_{r\rightarrow r_h}-\frac{1}{2}\sqrt{\frac{G^{rr}}{-\hat{G_{tt}}}}\hat{G_{tt}},r}{2 \pi}\Biggr|=\frac{r_h}{\sqrt{3} \pi ^{3/2}   \sqrt{N} \sqrt{g_s}}\left(\frac{1}{\gamma}+\frac{\beta}{2} \left(-{\cal C}_{zz}^{\rm BH} + 2 {\cal C}_{\theta_1z}^{\rm BH} - 3 {\cal C}_{\theta_1x}^{\rm BH}\right)\right),
\end{eqnarray}
where $\hat{G_{tt}}=-{\cal Y}_1(r)$, $\hat{G_{tt}},r$ implies derivative of $\hat{G_{tt}}$ with respect to $r$, and Lorentz factor, $\gamma=\frac{1}{\sqrt{1-l^2 \omega^2}}$. Since $L^4= 4 \pi g_s N$, therefore Hawking temperature of the rotating cylindrical black hole turns out to be:
\begin{eqnarray}
\label{Hawking-temp}
& &
\hskip -0.6in T_H(\gamma)= \left(\frac{r_h}{\pi L^2}\right)\left(\frac{1}{\gamma}+\frac{\beta}{2} \left(-{\cal C}_{zz}^{\rm BH} + 2 {\cal C}_{\theta_1z}^{\rm BH} - 3 {\cal C}_{\theta_1x}^{\rm BH}\right)\right) = T_H(0)\left(\frac{1}{\gamma}+\frac{\beta}{2}\left(-{\cal C}_{zz}^{\rm BH} + 2 {\cal C}_{\theta_1z}^{\rm BH} - 3 {\cal C}_{\theta_1x}^{\rm BH}\right)\right),
\end{eqnarray}
where $T_H(0)$ is the Hawking temperature of static black hole calculated in \cite{NPB}, ${\cal O}(\beta)$ term in the equations (\ref{Hawking-temp-defn}) and (\ref{Hawking-temp}) is calculated in small-$\omega$ approximation with $\gamma=1$ and ${\cal C}_{zz/ \theta_1z / \theta_1z}^{\rm BH}$ are integration constants which appear in the ${\cal O}(R^4)$ correction to the black hole background metric \cite{HD-Roorkee}. Therefore, Hawking temperature of the rotating cylindrical black hole will be scaled by the inverse of the Lorentz factor. Since, ${\cal C}_{zz}^{\rm BH}=2 {\cal C}_{\theta_1 z}^{\rm BH}$ and $|{\cal C}_{\theta_1 x}^{\rm BH}| \ll 1$ \cite{McTEQ}, therefore Hawking temperature recieves no ${\cal O}(R^4)$ correction, i.e.,
\begin{eqnarray}
\label{Hawking-temp-i}
& &
 T_H(\gamma)= \frac{T_H(0)}{\gamma}=T_H(0)\sqrt{1-l^2\omega^2}.
\end{eqnarray}
The metric of the ${\cal M}$-theory dual when we are restricting ourselves to the confined phase of thermal QCD-like theories, i.e., $T<T_c$ is  \cite{McTEQ}:
\begin{eqnarray}
\label{TypeIIA-from-M-theory-Witten-prescription-T<Tc}
\hskip -0.2in ds_{11}^2|_{Thermal} & = & e^{-\frac{2\phi^{\rm IIA}}{3}}\Biggl[\frac{1}{\sqrt{h(r,\theta_{1,2})}}\left(-dt^2 + \left(dx^1\right)^2 +  \left(dx^2\right)^2 + \left(dx^3\right)^2 \right)
\nonumber\\
& & \hskip -0.3in+ \sqrt{h(r,\theta_{1,2})}\left(dr^2 + ds^2_{\rm IIA}(r,\theta_{1,2},\phi_{1,2},\psi)\right)
\Biggr] + e^{\frac{4\phi^{\rm IIA}}{3}}\left(dx^{11} + A_{\rm IIA}^{F_1^{\rm IIB} + F_3^{\rm IIB} + F_5^{\rm IIB}}\right)^2.
\end{eqnarray}
To obtain the rotating cylindrical thermal background, we replace $x^3$ with $l \phi$ and perform the Lorentz transformation (\ref{Lorentz-boost}) of (\ref{TypeIIA-from-M-theory-Witten-prescription-T<Tc}) as we did for the black hole background, by doing so, metric of the rotating cylindrical thermal background (\ref{TypeIIA-from-M-theory-Witten-prescription-T<Tc}) is obtained  as
\begin{eqnarray}
\label{TypeIIA-from-M-theory-Witten-prescription-T<Tc-rotating-cylindrical}
\hskip -0.2in ds_{11}^2|_{Thermal} & = & e^{-\frac{2\phi^{\rm IIA}}{3}}\Biggl[\frac{1}{\sqrt{h(r,\theta_{1,2})}}\left(-dt^2 + \left(dx^1\right)^2 +  \left(dx^2\right)^2 +l^2\left(d\phi\right)^2 \right)
\nonumber\\
& & \hskip -0.4in+ \sqrt{h(r,\theta_{1,2})}\left(dr^2+ ds^2_{\rm IIA}(r,\theta_{1,2},\phi_{1,2},\psi)\right)
\Biggr] + e^{\frac{4\phi^{\rm IIA}}{3}}\left(dx^{11} + A_{\rm IIA}^{F_1^{\rm IIB} + F_3^{\rm IIB} + F_5^{\rm IIB}}\right)^2.
\end{eqnarray}

\section{Deconfinement Temperature at Intermediate Coupling in Rotating Plasma}
\label{Tc}
In this section, we will calculate the deconfinement temperature of rotating QGP at intermediate coupling from a semi-classical method \cite{Witten-Tc}. We need to equate the on-shell actions for the rotating cylindrical black hole and cylindrical thermal backgrounds at the UV cut-off for that purpose, i.e., $\beta_{\rm BH} S_{D=11, \rm on-shell}^{\rm BH}=\beta_{\rm th}S_{D=11, \rm on-shell}^{\rm th}$, where $\beta_{\rm BH}$ and $\beta_{\rm th}$ are periodicities of thermal circles in black hole and thermal backgrounds. \par

We have listed the metric for static black hole background including ${\cal O}(\beta)$ correction in \cite{MChPT} and metric of the thermal background (\ref{TypeIIA-from-M-theory-Witten-prescription-T<Tc-rotating-cylindrical}) is given in \cite{McTEQ}, the only difference is that $x^3=l\phi$ for this paper. In this paper, for black hole background metric along $(t,x^3)$ $(x^3=l\phi)$ subspace is different, and we have written those components in (\ref{metric-simp-t-phi}). The rest of the metric components are the same as \cite{MChPT}. Using the metric mentioned above, we calculated the various terms appearing in on-shell action (\ref{on-shell-D=11-action-up-to-beta}), e.g., Einstein-Hilbert and Gibbons-Hawking-York boundary terms. After substituting those terms in (\ref{on-shell-D=11-action-up-to-beta}), we integrated over angular coordinates $\left(\theta_{1,2},\phi_{1,2},\psi,x^{10}\right)$({eleven dimensional geometry is $S^1(t)\times_w \mathbb{R}^{1,2,3}(x^{1,2,3}) \times_w \mathbb{R}_{\geq 0}(r) \times_w {\cal M}_6(\theta_1,\theta_2,\phi_1,\phi_2,\psi,x^{10})$ where $\phi_{1,2} \in [0,2\pi]$, $\psi \in [0,4\pi]$ and $x^{10} \in [0,2\pi]$}). We found that dominant contributions to the ${\cal O}(\beta^{0})$ terms  $\left(S_{\rm EH}^{(0)}, S_{\rm GHY}^{(0)}\right)$ and ${\cal O}(\beta)$ terms $\left(S_{\rm EH}, \sqrt{-G_{\cal {M}}^{(1)}}R^{(0)} {\rm etc.}\right)$ in (\ref{on-shell-D=11-action-up-to-beta}) arise from the small $\theta_{1,2}$ values in terms of polar angular cut-offs $\epsilon_{1,2}$: $\theta_1\in\left[\epsilon_1,\pi-\epsilon_1\right]$ and $\theta_2\in\left[\epsilon_2,\pi-\epsilon_2\right]$. Further, radial integrals have been performed by partitioning $r$ into the IR ($r\in[r_{(h/0)},{\cal R}_{D5/\overline{D5}}^{\rm BH/th}]$) and the UV ($r\in[{\cal R}_{D5/\overline{D5}}^{\rm BH/th},{\cal R}_{\rm UV}]$). Explicitly, $\int_{r_{(h/0)}}^{{\cal R}_{\rm UV}} dr\left(S_{D=11}^{\rm on-shell, \ (BH/th)}\right) =\int_{r_{(h/0)}}^{{\cal R}_{D5/\overline{D5}}^{\rm BH/th}} dr\left(S_{D=11}^{\rm on-shell, \ (BH/th)}\right) +\int_{{\cal R}_{D5/\overline{D5}}^{\rm BH/th}}^{{\cal R}_{\rm UV}} dr\left(S_{D=11}^{\rm on-shell, \ (BH/th)}\right)$, where, $r_h$ is the non-extremality parameter when $T>T_C$ and $r_0$ is the IR cut off when $T<T_c$, $``BH"$ and $``th"$ implying black hole and thermal backgrounds. Radial coordinate everywhere in this paper and in \cite{MChPT, McTEQ, HD-Roorkee} is taken as $r \equiv \frac{r}{{\cal R}_{D5/\overline{D5}}}$ \cite{Bulk-viscosity}.
In the end, we obtain on-shell action densities and are writing the final results.

\subsection{Deconfinement Temperature of Rotating QGP from ${\cal M}$-Theory Dual}
 \label{O-beta0-Tc}
\begin{figure}
\begin{center}
\includegraphics[width=0.60\textwidth]{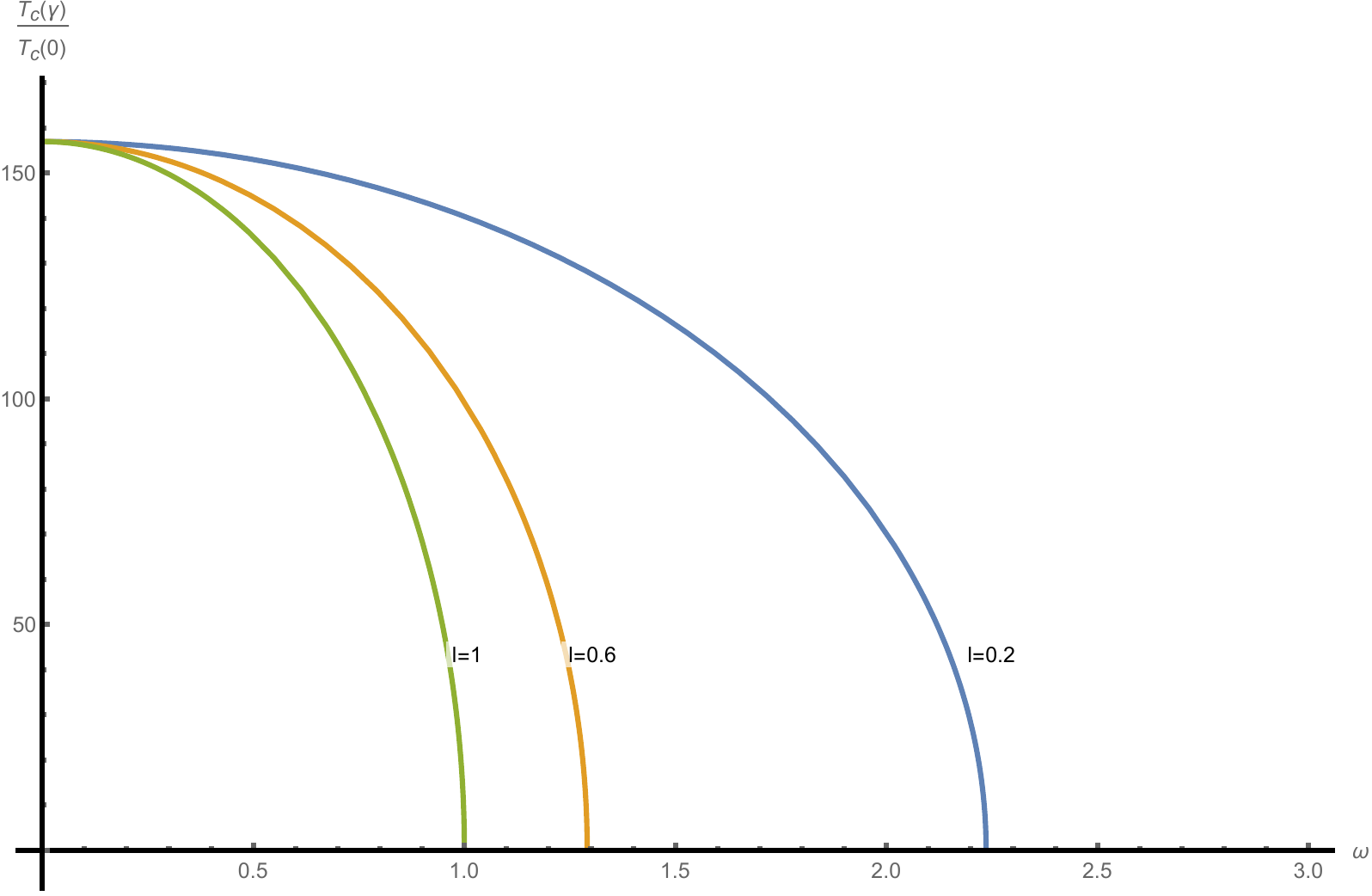}
\end{center}
\caption{Plot of $\frac{T_c(\gamma)}{T_c(0)}$ versus $\omega$ for $l=0.2,0.6,1$.}
\label{Tc-Plot}
\end{figure}

 By following the procedure discussed above, at ${\cal O}(\beta^0)$, UV-finite and holographic IR regularised on-shell action density for the ${\cal M}$-theory rotating cylindrical black hole background uplift (\ref{TypeIIA-from-M-theory-Witten-prescription-T>Tc-Rotation-Canonical}) is obtained as
{\footnotesize
\begin{eqnarray}
\label{on-shell-BH-rotation}
& & \hskip -0.7in \left(1+\frac{r_h^4}{2{\cal R}_{\rm UV}^4}\right)\frac{S_{D=11,\ {\rm on-shell\ UV-finite}}^{\rm BH}}{{\cal V}_4} = \lambda_{\rm EH, IR}^{{\rm BH}} \frac{ \epsilon^{\rm BH}  \gamma ^8 \omega^2 M N_f^3 g_s^{3/2} r_h^4 \log ^3(N) \log \left(\frac{r_h}{{\cal R}_{D5/\overline{D5}}^{\rm BH}}\right) \log \left(1-\frac{r_h}{{\cal R}_{D5/\overline{D5}}^{\rm BH}}\right)}{{{\cal R}_{D5/\overline{D5}}^{\rm BH}}^4N^{1/2}} \nonumber\\
& &\hskip -0.7in + \lambda_{\rm EH, UV}^{{\rm BH}} \frac{ \gamma ^8 l M^{\rm UV}   r_h^4  \log ^2\left(\frac{{\cal R}_{\rm UV}}{{\cal R}_{D5/\overline{D5}}^{\rm BH}}\right)}{N^{1/2}{{g_s^{\rm UV}}^{3/2}} {{\cal R}_{D5/\overline{D5}}^{\rm BH}}^4 }  + \lambda_{\rm GHY}^{{\rm BH}}\frac{l M^{\rm UV}  r_h^4  \log\left(\frac{{\cal R}_{\rm UV}}{{\cal R}_{D5/\overline{D5}}^{\rm BH}}\right)}{N^{1/2} {{\cal R}_{D5/\overline{D5}}^{\rm BH}}^4  {{g_s^{\rm UV}}^{3/2}}},
\end{eqnarray}
}
where $\lambda_{\rm EH, IR}^{{\rm BH}}, \ \lambda_{\rm EH, UV}^{\rm BH}$
and $\lambda_{\rm GHY}^{{\rm BH}}$ are the numerical prefactors.  ${\cal V}_4$ is the coordinate volume of $S^1(t)\times_w \mathbb{R}^2(x^{1,2})\times_w S^1(\phi)$ and ${\cal R}_{D5-\overline{D5}}^{\rm {BH/th}}\equiv \sqrt{3}a^{\rm {BH/th}}$, where $a^{\rm {BH/th}}$ are the resolution parameter of the blown up $S^2$ and are defined as $a^{\rm {BH/th}} = \left(\frac{1}{\sqrt{3}} + \epsilon^{\rm {BH/th}} + {\cal O}\left(\frac{g_sM^2}{N}\right)\right)r_{(h/0)}$ \cite{HD-Roorkee}. Further, UV-finite and holographic IR regularized ${\cal O}(\beta^0)$ contribution of the on-shell action density for the ${\cal M}$-theory rotating cylindrical thermal background uplift (\ref{TypeIIA-from-M-theory-Witten-prescription-T<Tc-rotating-cylindrical}) turns out be:
{\footnotesize
\begin{eqnarray}
\label{on-shell-th-rotation}
& & \frac{S_{D=11,\ {\rm on-shell\ UV-finite}}^{\rm thermal}}{{\cal V}_4} =  \frac{ \lambda_{\rm GHY}^{{\rm th}} {M_{\rm UV}}{\it l} {r_0}^4 \log \left(\frac{{{\cal R}_{\rm UV}}}{{\cal R}_{D5/\overline{D5}}^{\rm th}}\right)}{{{g_s^{\rm UV}}^{3/2}}
   N^{1/2} {{\cal R}_{D5/\overline{D5}}^{\rm th}}^4 }+\frac{{g_s}^{3/2} \lambda_{\rm EH, IR}^{{\rm th}} M {N_f}^3{\it l} {r_0}^2 \log^2(N) \log \left(\frac{{r_0}}{{\cal R}_{D5/\overline{D5}}^{\rm th}}\right)}{ N^{1/2} {{\cal R}_{D5/\overline{D5}}^{\rm th}}^2 }\nonumber\\
& &  + \frac{\lambda_{\rm EH,\ UV}^{\rm th}{M_{\rm UV}} {N_f^{\rm UV}}{\it l}  \left(-\frac{121 {r_0}^4}{16 {{\cal R}_{D5/\overline{D5}}^{\rm th}}^4}-\frac{6 {r_0}^2}{{{\cal R}_{D5/\overline{D5}}^{\rm th}}^2}+2\right)}{ {g_s^{\rm UV}}^{1/2} N^{\frac{1}{2}} }, 
\end{eqnarray}  
}
where $r_0$ is the IR cut-off of the theory when $T<T_c$ in QCD and $\lambda_{\rm GHY}^{{\rm th}}, \ \lambda_{\rm EH, IR}^{{\rm th}}$ and $\lambda_{\rm EH,\ UV}^{\rm th}$ are the numerical prefactors. Parameters with superscript ``${\rm UV}$'' like $M^{\rm UV}, g_s^{\rm UV}$ etc. in (\ref{on-shell-BH-rotation}) and (\ref{on-shell-th-rotation}) are the UV-valued parameters. At the UV-cutoff \cite{McTEQ}:
\begin{equation}
\label{SBH=Sth}
\left(1+\frac{r_h^4}{2{\cal R}_{\rm UV}^4}\right)\frac{S_{D=11,\ {\rm on-shell\ UV-finite}}^{\rm BH}}{{\cal V}_4}=\frac{S_{D=11,\ {\rm on-shell\ UV-finite}}^{\rm thermal}}{{\cal V}_4}.
\end{equation}
In equation (\ref{on-shell-BH-rotation}), since $\omega^2<1$, $\lambda_{\rm GHY}^{{\rm BH}} \sim {\cal O}(10^3)\lambda_{\rm EH, IR}^{{\rm BH}} $  and $\lambda_{\rm GHY}^{{\rm BH}} \sim {\cal O}(10)\lambda_{\rm EH, UV}^{{\rm BH}} $, and in  equation (\ref{on-shell-th-rotation}), $\lambda_{\rm GHY}^{{\rm th}} \sim {\cal O}(10^3)\lambda_{\rm EH, IR}^{{\rm th}} $, $\lambda_{\rm GHY}^{{\rm th}} \sim {\cal O}(10^2)\lambda_{\rm EH,\ UV}^{\rm th}$, therefore, one is required to solve the following equation:
\begin{eqnarray}
\label{rh-r0-equation}
& &  \lambda_{\rm GHY}^{{\rm BH}}\frac{l M^{\rm UV}  r_h^4  \log\left(\frac{{\cal R}_{\rm UV}}{{\cal R}_{D5/\overline{D5}}^{\rm BH}}\right)}{N^{1/2} {{\cal R}_{D5/\overline{D5}}^{\rm BH}}^4  {{g_s^{\rm UV}}^{3/2}}}-\frac{ \lambda_{\rm GHY}^{{\rm th}} {M_{\rm UV}}{\it l} {r_0}^4 \log \left(\frac{{{\cal R}_{\rm UV}}}{{\cal R}_{D5/\overline{D5}}^{\rm th}}\right)}{{{g_s^{\rm UV}}^{3/2}} N^{1/2} {{\cal R}_{D5/\overline{D5}}^{\rm th}}^4 }=0.
\end{eqnarray}  
Solution to the above equation is: 
\begin{eqnarray}
  \label{rh-r0-rotation-beta0}  
 & &r_h= \frac{\sqrt[4]{\frac{ \lambda_{\rm GHY}^{{\rm th}}}{\lambda_{\rm GHY}^{{\rm BH}}}} r_0 {{\cal R}_{D5/\overline{D5}}^{\rm BH}} \sqrt[4]{\frac{\log
   \left(\frac{{{\cal R}_{\rm UV}}}{{{\cal R}_{D5/\overline{D5}}^{\rm th}}}\right)}{\log
   \left(\frac{{{\cal R}_{\rm UV}}}{{{\cal R}_{D5/\overline{D5}}^{\rm BH}}}\right)}}}{{\cal R}_{D5/\overline{D5}}^{\rm th}}. 
\end{eqnarray}
Using the numerical values: $\lambda_{\rm GHY}^{{\rm th}}=0.08$, $\lambda_{\rm GHY}^{{\rm BH}}=0.23$; Since, ${{\cal R}_{D5/\overline{D5}}^{\rm th/BH}}=\sqrt{3}\left(\frac{1}{\sqrt{3}}+\epsilon^{\rm th/BH} \right)e^{-\kappa_{(h/0)} N^{-1/3}}$, where $\kappa_{(h/0)}\ll 1$ \cite{Bulk-viscosity}, ${\cal R}_{\rm UV} \approx L=(4 \pi g_s N)^{1/4}$ \cite{MChPT} and choosing $g_s=0.1$, $N=100$, $\kappa_{r_{h}}=0.01$, $\kappa_{r_{0}}=0.01$, $\epsilon^{\rm BH}=0.36$ and $\epsilon^{\rm th}=0.1$, equation (\ref{rh-r0-rotation-beta0}) simplifies to
\begin{eqnarray}
\label{rh-r0-simp}
& & r_h \approx 1.16 r_0.
\end{eqnarray}
In gauge-gravity duality, Hawking Page phase transition between a thermal and black hole background at $T=T_H(\gamma)$ on the gravity side is dual to the deconfinement phase transition between confined phase to deconfined phase of gauge theories at $T=T_c(\gamma)$, i.e., $T_H(\gamma) \sim T_c(\gamma)$ \cite{Witten-Tc}. Therefore, using (\ref{Hawking-temp-i}) and (\ref{rh-r0-simp}), the deconfinement temperature of the thermal QCD-like theories in the presence of rotating Quark Gluon Plasma is obtained as:
\begin{eqnarray}
\label{Tc-rotation}
& & \hskip -0.5in
T_c(\gamma)=\frac{1}{\gamma}\Biggl( \frac{r_h}{\pi L^2}\Biggr)=\frac{1}{\gamma}\Biggl( \frac{1.16 r_0}{\pi \sqrt{4 \pi g_s N}}\Biggr)=T_c(0)\sqrt{1-l^2 \omega^2}. 
\end{eqnarray}
Using $\frac{r_0}{\sqrt{4 \pi g_s N}} = \frac{1700}{4}$MeV \cite{Misra+Gale}, we obtained $T_c(0) \approx 157$ MeV, which is close to the value obtained from lattice calculations \cite{Tc-L} and holographic study \cite{Tc-HW_SW}. For $l=0.2,0.6,1 \ GeV^{-1 }$ plot of the ratio of deconfinement temperature of QCD in the presence and absence of rotation versus angular velocity of the rotating quark-gluon plasma is given in Fig. \ref{Tc-Plot} which is justifying equation (\ref{Tc-rotation}). From Fig. \ref{Tc-Plot}, we see that as the angular velocity of rotating QGP increases, $T_c$ decreases and vice-versa. Further, when $l$ decreases, a large angular velocity is required to reduce the deconfinement temperature.\\
{\bf Comparison with Other Models}: The result that decrease of deconfinement temperature of thermal QCD with the increase of rotation of quark-gluon plasma was earlier studied using gauge-gravity duality from a bottom-up approach in \cite{R-Tc,G-D-Rotation,Kerr-AdS-1,Kerr-AdS-2}, authors observed same behavior from Nambu-Jona-Lansinio model in \cite{R-Tc3} because of chiral condensate suppression, this was also studied from lattice simulations in \cite{Lattice-1,Lattice-2}. Lattice simulations have been performed in rotating reference frames with various boundary conditions (open, periodic, and Dirichlet). The authors therein studied the effect of rotation on the critical temperature in gluodynamics and in the theory of dynamical fermions. In the case of gludynamics, critical temperature exhibits opposite behavior compared to our result and other holographic results \cite{R-Tc,G-D-Rotation,Kerr-AdS-1,Kerr-AdS-2} while for the dynamical fermions, the authors observed the same behaviour as ours, i.e., critical temperature decreases with the increase of rotation. The bottom-up holographic study of \cite{R-Tc} was performed using hard-wall and soft-wall models, gravity dual of \cite{G-D-Rotation} involves Einstein-Maxwell-Dilaton system; authors in \cite{Kerr-AdS-1,Kerr-AdS-2} considered gravity dual of rotating QGP as Kerr-AdS black hole background in five dimensions. No known paper discusses the deconfinement temperature of rotating QGP from a top-down model. Top-down models are more fundamental because one starts with original type IIB/IIA string theory or ${\cal M}$-theory and then compactify the same on an internal manifold to obtain four-dimensional QCD at finite temperature. The model that we are using in this paper is very close to QCD because this involves the finite number of color and flavors in QCD, and exhibits IR confinement as well as UV conformality. For the first time, we successfully studied rotating QGP from a top-down approach and obtained a result consistent with phenomenological and lattice results, which confirms the non-zero angular velocity of rotating QGP obtained from hydrodynamic simulation \cite{omega-simulation} and STAR collaboration \cite{RHIC-omega}.

\subsection{UV-IR Mixing, Flavor Memory Effect and Non-Renormalization of $T_c$ in Rotating QGP}
\label{UV-IR-mixing}
Quoting the results from appendix \ref{beta}. UV finite on-shell action densities for the rotating cylindrical black hole and thermal backgrounds at ${\cal O}(\beta)$ are
{\footnotesize
\begin{eqnarray}
\label{bh-th-action-beta}
& & \left(1+\frac{r_h^4}{2{\cal R}_{\rm UV}^4}\right)\frac{S_{D=11,\ {\rm on-shell \ UV-finite}}^{\beta, \rm BH}}{{\cal V}_4} = \Biggl[{-2 {\cal C}_{\theta_1x} \kappa_{\sqrt{G^{(1)}}R^{(0)}}^{\rm IR} } +\frac{20  \left(-{\cal C}_{zz}^{\rm bh} + 2 {\cal C}_{\theta_1z}^{\rm bh} - 3 {\cal C}_{\theta_1x}^{\rm bh}\right) \kappa_{\rm EH}^{\beta,\ \rm IR} }{11 }\Biggr]\nonumber\\
& & \times\frac{ b^2   {g_s}^{3/2}  M {N_f}^3 {r_h}^4 \log ^3(N) \log
   \left(\frac{{r_h}}{{{\cal R}_{D5/\overline{D5}}}}\right) \log
   \left(1 - \frac{{r_h}}{{{\cal R}_{D5/\overline{D5}}}}\right)}{ N^{1/2} {{\cal R}_{D5/\overline{D5}}}^4}\beta,\nonumber\\
   & & \frac{S_{D=11,\ {\rm on-shell \ UV-finite}}^{\beta,\rm thermal}}{{\cal V}_4} =
 -\frac{20  \kappa_{\rm EH, th}^{{\rm IR},\ \beta} {r_0}^3 N^{1/2}{f_{x^{10}x^{10}}}({r_0})}{11{g_s}^{3/2} M  {N_f}^{5/3} {{\cal R}_{D5/\overline{D5}}^{\rm th}}^3  \log ^{\frac{2}{3}}(N) \log
   \left(\frac{{r_0}}{{\cal R}_{D5/\overline{D5}}^{\rm th}}\right)}\beta.
\end{eqnarray}
}
Similar to the ${\cal O}(\beta^0)$ case (\ref{SBH=Sth}), equating on-shell action density contributions from ${\cal O}(\beta)$ terms appearing in (\ref{bh-th-action-beta}) at the UV cut-off, i.e., 
\begin{equation}
\label{SBH=Sth-i}
\left(1+\frac{r_h^4}{2{\cal R}_{\rm UV}^4}\right)\frac{S_{D=11,\ {\rm on-shell\ UV-finite}}^{\beta, \rm BH}}{{\cal V}_4}=\frac{S_{D=11,\ {\rm on-shell\ UV-finite}}^{\beta, \rm thermal}}{{\cal V}_4}. 
\end{equation}
From equations (\ref{bh-th-action-beta}) and (\ref{SBH=Sth-i}), we obtained
{\footnotesize
\begin{eqnarray}
\label{UV-IR mixing-rotation}
& & \hskip -0.5in f_{x^{10}x^{10}}({r_0})= -\frac{b^2 {g_s}^3 M^2 {N_f}^{14/3}  \left(\frac{r_h}{{\cal R}_{D5/\overline{D5}}^{\rm bh}}\right)^4  \log ^{{3}}(N) \log
   \left(\frac{{r_0}}{{\cal R}_{D5/\overline{D5}}^{\rm th}}\right) \log \left(\frac{{r_h}}{{\cal R}_{D5/\overline{D5}}^{\rm bh}}\right) \log
   \left(1-\frac{{r_h}}{{\cal R}_{D5/\overline{D5}}^{\rm bh}}\right)}{  {\kappa_{\rm EH, th}^{\beta,\ \rm IR}} {N} \left(\frac{r_0}{{\cal R}_{D5/\overline{D5}}^{\rm th}}\right)^3}\nonumber\\
& & \hskip 0.3in \times \left(-11  {\cal C}_{\theta_1x} {\kappa_{\sqrt{G^{(1)}}R^{(0)}}^{\rm IR}}
    \log ^3(N)-10 {\kappa_{\rm EH,bh}^{\beta,\ \rm IR}} \left(-{\cal C}_{zz}^{\rm bh} + 2 {\cal C}_{\theta_1z}^{\rm bh} - 3 {\cal C}_{\theta_1x}^{\rm bh}\right)\right).
\end{eqnarray}
}
Equation (\ref{UV-IR mixing-rotation}) is a version of ``UV-IR'' mixing similar to that observed in \cite{McTEQ}. Combination of integration constants, $\left(-{\cal C}_{zz}^{\rm bh} + 2 {\cal C}_{\theta_1z}^{\rm bh} - 3 {\cal C}_{\theta_1x}^{\rm bh}\right)$, has the memory of flavor $D7$-branes of type IIB string dual known as ``Flavor Memory'' effect, and the same appeared in ${\cal O}(\beta)$ contribution to the Hawking temperature (\ref{Hawking-temp}). In subsection \ref{Rotating-QGP}, we have discussed that the aforementioned combination of constants of integration constant is zero, which implies no ${\cal O}(\beta)$ correction to Hawking temperature (\ref{Hawking-temp-i}) and hence no ${\cal O}(\beta)$ correction to the deconfinement temperature $(T_c)$. This implies ``non-renormalization of $T_c$'' at ${\cal O}(\beta)$. Finally, one can show that for a rotating cylindrical black hole background, UV divergences appearing at ${\cal O}(\beta^0)$ can be canceled by boundary Einstein-Hilbert term $\left(\int_{r={\cal R}_{\rm UV}} \sqrt{-h^{\cal M}}R\right)$ and for thermal background, one needs two boundary counter terms which are boundary cosmological constant term $\left(\int_{r={\cal R}_{\rm UV}} \sqrt{-h^{\cal M}}\right)$ and boundary Einstein-Hilbert term $\left(\int_{r={\cal R}_{\rm UV}} \sqrt{-h^{\cal M}} R\right)$. Further, UV divergence appearing at  ${\cal O}(\beta)$ in the on-shell action (\ref{on-shell-D=11-action-up-to-beta}) coming from  ${\sqrt{-G^{\cal M}}G^{MN}\frac{\delta J_0}{\delta G^{MN}}}$ will be canceled by ${\int_{r={\cal R}_{\rm UV}}\sqrt{-h^{\cal M}}G^{mn}\frac{\delta J_0}{\delta G^{mn}}}$ for both rotating cylindrical black hole and thermal backgrounds with some constraint on the number of flavor branes in the UV $(N_f^{\rm UV})$.

\section{Conclusion}
\label{summary}
With the aim of studying deconfinement phase transition in rotating Quark Gluon Plasma {\it at finite coupling} via gauge-gravity duality involving a Hawking-Page phase transition on the gravity dual side from the thermal background ($T<T_c$ on gauge theory side) to the black hole background ($T>T_c$ on gauge theory side), we have calculated the deconfinement temperature of rotating quark-gluon plasma from the ${\cal M}$-theory dual inclusive of ${\cal O}(R^4)$ correction \cite{HD-Roorkee} from a semi-classical method advocated in \cite{Witten-Tc}.   \par
 
\par
 In this paper, we studied the effect of rotation on the deconfinement temperature of thermal QCD-like theories by computing the various terms appearing in the on-shell action densities of the rotating cylindrical black hole and thermal backgrounds and then equated these two at the UV cut-off. {\it Using gauge-gravity duality, we obtain the rotating QGP on the gauge theory side by performing Lorentz transformations (\ref{Lorentz-boost}) along $(t,\phi)$ coordinates on the gravity dual side. Hence, gravity dual involves rotating cylindrical black hole and thermal backgrounds when $T>T_c$ and $T<T_c$ on the gauge theory side, where $T_c$ is the deconfinement temperature of thermal QCD-like theories}. On equating the ${\cal O}(\beta^0)$ terms for the rotating cylindrical black hole and cylindrical thermal backgrounds at the UV cut-off (\ref{SBH=Sth}), we obtained the deconfinement temperature of rotating QGP (\ref{Tc-rotation}) which is inversely proportional to the Lorentz factor, $\gamma$. This implies that {\it deconfinement temperature of thermal QCD-like theories decrease when we increase the angular velocity of rotating QGP}. The same effect that the deconfinement temperature of QCD decreases with the increase of rotation was also discussed in \cite{Maxim}. Other papers in the literature show the same behavior of the deconfinement temperature of rotating QGP that we obtained in this paper, e.g., see \cite{R-Tc,R-Tc-1}. When we are equating the ${\cal O}(\beta)$ terms in the on-shell actions of the rotating cylindrical black hole and thermal backgrounds at the UV cut-off, we observed a novel {\it ``UV-IR'' mixing} (we equated the on-shell actions at the UV cut-off and we obtained a relationship in the IR between the combination of constants of integrations appearing in ${\cal O}(R^4)$ correction to the metric of the rotating cylindrical black hole background and ${\cal O}(R^4)$ correction to the thermal background along the ${\cal M}$-theory circle (\ref{UV-IR mixing-rotation})) in our setup. Even though there are no branes in ${\cal M}$-theory uplift, the aforementioned combination of constants of integrations appearing in ${\cal O}(R^4)$ correction to the metric of the black hole background are precisely along the compact part $(S^3(\theta_1,x(\phi_1),z(\psi_1)))$ of the non-compact four cycle $(\mathbb{R}_+ \times S^3(\theta_1,x(\phi_1),z(\psi_1))$ wrapped by the flavor $D7$ branes in the parent type IIB dual \cite{metrics}. Therefore, this retains the ``memory'' of flavor $D7$ branes of parent type IIB dual \cite{metrics} and is therefore interpreted as {\it ``Flavor Memory'' effect} in the ${\cal M}$-theory dual. We have calculated the Hawking temperature of rotating cylindrical black hole background in (\ref{Hawking-temp}), which is proportional to $\left(-{\cal C}_{zz}^{\rm BH} + 2 {\cal C}_{\theta_1z}^{\rm BH} - 3 {\cal C}_{\theta_1x}^{\rm BH}\right)$. It was shown in \cite{McTEQ} that ${\cal C}_{zz}^{\rm BH}=2 {\cal C}_{\theta_1 z}^{\rm BH}$ and $|{\cal C}_{\theta_1 x}^{\rm BH}| \ll 1$, to set the correction to the three form potential zero. This implies that there are no ${\cal O}(R^4)$ correction to the deconfinement temperature even in the presence of rotating Quark Gluon Plasma, and this indicates the  {\it ``Non-Renormalization of the deconfinement temperature''} in the rotating QGP as well. From the above discussion, we find that, {\it ``UV-IR'' mixing, ``Flavor Memory'' effect and ``Non-Renormalization of $T_c$'' exists even in the rotating QGP}. Further, we have regularised the action by adding appropriate counter terms to cancel the UV divergences for rotating cylindrical black hole, and thermal backgrounds and IR divergences have been removed by doing holographic IR regularization.

\par Experimentally, it was found that Quark Gluon Plasma produced in noncentral Relativistic Heavy Ion Collider (RHIC) has angular velocity and deconfinement temperature decreases with the increase of rotation \cite{RHIC-omega}. Although this has thus far been verified using gauge-gravity duality from a bottom-up approach \cite{R-Tc,G-D-Rotation,Kerr-AdS-1,Kerr-AdS-2}, our work is a good check for the top-down duals of \cite{metrics}, \cite{HD-Roorkee} in the context of the computation of deconfinement temperature of rotating QGP from a top-down approach consistent with the experimental results.

\vspace{-0.3in}
\section*{Acknowledgements}
\vspace{-0.1in}
GY is supported by a Senior Research Fellowship (SRF) from the Council of Scientific and Industrial Research, Govt. of India. I would like to thank A.~Misra for suggesting this problem to me and very helpful discussions during this project. I would also like to thank V.~Yadav for useful clarifications. I am also thankful to Y.~Yadav for wonderful discussions.
\appendix
\section{${\cal M}$-Theory Metric}
\label{metric}
\setcounter{equation}{0}
\seceqaa
\begin{itemize}
\item The following ${\cal O}(\beta^0)$ components of the ${\cal M}$-theory metric involving a rotating cylindrical black hole in the MQGP limit are different from ones as obtained in \cite{MQGP,NPB}.
{\scriptsize
\begin{eqnarray}
\label{metric-simp-t-phi}
& & 
 G^{\cal M}_{tt}(r)= \frac{\left(\frac{1}{N}\right)^{3/2} r^2 \left(a_1(r) (N-B(r))-a_2(r)\right)}{2 \sqrt{\pi } \sqrt[3]{a_1(r)} \sqrt{g_s}}\left(\frac{g(r)-l^2\omega^2}{1-l^2\omega^2}\right), \nonumber\\
 & & G^{\cal M}_{t \phi}(r)= G^{\cal M}_{\phi t}(r)=\frac{\left(\frac{1}{N}\right)^{3/2} r^2  \left(a_1(r) (N-B(r))-a_2(r)\right)}{2 \sqrt{\pi } \sqrt[3]{a_1(r)} \sqrt{g_s}}\left(\frac{l^2\omega(1-g(r))}{1-l^2\omega^2}\right),\nonumber\\
& & G^{\cal M}_{\phi \phi}(r)=\frac{\left(\frac{1}{N}\right)^{3/2} r^2 \left(a_1(r) (N-B(r))-a_2(r)\right)}{2 \sqrt{\pi } \sqrt[3]{a_1(r)} \sqrt{g_s}}\left(\frac{-g(r)l^4\omega^2+l^2}{1-l^2\omega^2}\right),
\end{eqnarray}
}
where
{\scriptsize
\begin{eqnarray}
& & 
a_1(r)=\frac{3 \left(-N_f \log \left(9 a^2 r^4+r^6\right)-4 N_f \log \left(\alpha _{\theta _1}\right)-4 N_f \log \left(\alpha _{\theta _2}\right)+2 N_f \log (N)+4 \log (4) N_f+\frac{8 \pi }{g_s}\right)}{8 \pi },\nonumber\\
& & a_2(r)=\frac{12 a^2 M^2 N_f g_s \left(c_2 \log \left(r_h\right)+c_1\right)}{9 a^2+r^2}; \ \  a_3(r)=\frac{r^2 a_2(r)}{2 {N_f} \left(6 a^2+r^2\right)}; \ \ a_4(r)=\frac{6 a^2+r^2}{9 a^2+r^2},\nonumber\\
& & b_1(r)=\frac{M g_s^{3/4}}{6 \sqrt{2} \pi ^{5/4} r^2 \alpha _{\theta _1} \alpha _{\theta _2}^2}\Biggl[\left(\log (r) \left(4 \left(r^2-3 a^2\right) N_f g_s \log \left(\frac{1}{4} \alpha _{\theta _1} \alpha _{\theta _2}\right)+8 \pi  \left(r^2-3 a^2\right)-3 {g_s} r^2 N_f\right) \right)\nonumber\\
& & \  +\left( \left(r^2-3 a^2\right) N_f g_s \log \left(\frac{1}{4} \alpha _{\theta _1} \alpha _{\theta _2}\right)+18 \left(r^2-3 a^2 (6 r+1)\right) N_f g_s \log ^2(r)-\left(r^2-3 a^2\right) N_f g_s \log (N) (2\log (r)+1)\right)\Biggr],\nonumber\\
& & b_2(r)=\frac{M N_f g_s^{7/4} \log (r) \left(36 a^2 \log (r)+r\right)}{3 \sqrt{2} \pi ^{5/4} r \alpha _{\theta _2}^3}; \  B_1(r)=\frac{2 \sqrt{\pi } \sqrt{N} \sqrt{g_s}}{r^2} ; g(r)=1-\frac{r_h^4}{r^4}\ ,\nonumber\\
& & B(r)=\frac{3 g_s M^2 \log (r)}{32 \pi ^2}\times \Biggl[\left(12 N_f g_s \log (r)+2 N_f g_s \log \left(\alpha _{\theta _1}\right)+2 N_f g_s \log \left(\alpha _{\theta _2}\right)+6 N_f g_s\right)   +\left(N_f g_s (-\log (N))-2 \log (4) N_f g_s+8\pi \right)\Biggr].\nonumber\\
\end{eqnarray}
}
\item The ${\cal O}(\beta)$ terms in the small $\omega$-limit remain the same as worked out in \cite{HD-Roorkee}.

\item ${\cal M}$-theory metric involving a rotating cylindrical thermal gravitational background in the MQGP limit can be obtained from the ${\cal M}$-theory metric involving a thermal background given in \cite{McTEQ} just by replacing $x^3$ by $ l \phi$ in \cite{McTEQ}.
\end{itemize}

\section{${\cal O}(\beta)$ Contribution to the On-Shell Action Densities}
\setcounter{equation}{0}
\seceqbb
\label{beta}
To obtain the contributions from the ${\cal O}(\beta)$ terms in the on-shell action density for the black hole background, one is required to write the metric (\ref{TypeIIA-from-M-theory-Witten-prescription-T>Tc-Rotation-Canonical}) in diagonal basis. Unwarped metric in the $t-\phi$ subspace can be rewritten as
\begin{eqnarray}
\label{metric-tphi-diagonal}
ds^2=-\left(1-\frac{r_h^4}{r^4 \gamma^2}\right)dT^2+l^2 d\Phi^2,
\end{eqnarray}
where
\begin{eqnarray}
\label{coordinate-transformations}
& & 
dT=dt - \frac{l^2 r_h^4 \omega d\phi}{r^4-r_h^4}; \ d\Phi=\frac{l^2 r_h^4 \omega dt}{r^4-r_h^4}+ d\phi .
\end{eqnarray}
The ${\cal O}(\beta)$ corrected metric in the diagonal basis can be written as
\begin{eqnarray}
G_{MN}^{\cal M}=G_{MN}^{\rm MQGP}\left(1+\beta f_{MN}(r)\right),
\end{eqnarray}
where $f_{MN}(r)$ are given in \cite{HD-Roorkee}. Given that 
we are working at ${\cal O}(\beta)$, we drop terms of ${\cal O}(\omega^2)$ in (\ref{coordinate-transformations})-(\ref{metric-tphi-diagonal-small-l-omega}). At ${\cal O}(\beta)$, finite contributions to the on-shell action densities for the rotating cylindrical black hole and thermal backgrounds have been obtained using a similar procedure (summarised before starting of subsection \ref{O-beta0-Tc}) as done in the calculation of ${\cal O}(\beta^0)$ contributions(subsection \ref{O-beta0-Tc}) and the final results are\footnote{In the small-$\omega$ limit, calculations at ${\cal O}(\beta)$ are almost similar to \cite{McTEQ}.}
\begin{eqnarray}
\label{bh-action-beta}
& & \left(1+\frac{r_h^4}{2{\cal R}_{\rm UV}^4}\right)\frac{S_{D=11,\ {\rm on-shell \ UV-finite}}^{\beta, \rm BH}}{{\cal V}_4} = \Biggl[{-2 {\cal C}_{\theta_1x} \kappa_{\sqrt{G^{(1)}}R^{(0)}}^{\rm IR} } +\frac{20  \left(-{\cal C}_{zz}^{\rm bh} + 2 {\cal C}_{\theta_1z}^{\rm bh} - 3 {\cal C}_{\theta_1x}^{\rm bh}\right) \kappa_{\rm EH}^{\beta,\ \rm IR} }{11 }\Biggr]\nonumber\\
& & \times\frac{ b^2   {g_s}^{3/2}  M {N_f}^3 {r_h}^4 \log ^3(N) \log
   \left(\frac{{r_h}}{{{\cal R}_{D5/\overline{D5}}}}\right) \log
   \left(1 - \frac{{r_h}}{{{\cal R}_{D5/\overline{D5}}}}\right)}{ N^{1/2} {{\cal R}_{D5/\overline{D5}}}^4}\beta,
\end{eqnarray}
and
\begin{eqnarray}
\label{th-action-beta}
& & \frac{S_{D=11,\ {\rm on-shell \ UV-finite}}^{\beta,\rm thermal}}{{\cal V}_4} =
 -\frac{20  \kappa_{\rm EH, th}^{{\rm IR},\ \beta} {r_0}^3 N^{1/2}{f_{x^{10}x^{10}}}({r_0})}{11{g_s}^{3/2} M  {N_f}^{5/3} {{\cal R}_{D5/\overline{D5}}^{\rm th}}^3  \log ^{\frac{2}{3}}(N) \log
   \left(\frac{{r_0}}{{\cal R}_{D5/\overline{D5}}^{\rm th}}\right)}\beta.  
\end{eqnarray}  
Now using equation (\ref{coordinate-transformations}), we can write the metric in $(t,\phi)$ subspace and ${\cal O}(R^4)$ correction to the same, i.e., $f_{MN}(r)$, where $(M,N=t,\phi)$ (we are focusing only on the $t,\phi$ subspace). We can write the $tt$ component of the metric as
\begin{eqnarray}
& & 
g_{tt}=\left(\frac{dT}{dt}\right)^2 G_{TT}^{\cal M}+\left(\frac{d\Phi}{d\phi}\right)^2 G_{\Phi \Phi}^{\cal M},
\end{eqnarray}
the above equation implies that
\begin{eqnarray}
& & f_{tt}=\left(\frac{dT}{dt}\right)^2 f_{TT}+\left(\frac{d\Phi}{d\phi}\right)^2 f_{\Phi \Phi}= f_{TT}+ f_{\Phi \Phi}.
\end{eqnarray}
Similarly, we can write $\phi t$ component of the metric as:
\begin{eqnarray}
& & g_{\phi t}=\left(\frac{dT}{d\phi}\right)\left(\frac{dT}{dt}\right) G_{TT}^{\cal M}+\left(\frac{dT}{d\phi}\right)\left(\frac{dT}{dt}\right) G_{\Phi \Phi}^{\cal M},
\end{eqnarray}
and hence 
\begin{eqnarray}
& & f_{\phi t}=\left(\frac{dT}{d\phi}\right)\left(\frac{dT}{dt}\right) f_{TT}+\left(\frac{dT}{d\phi}\right)\left(\frac{dT}{dt}\right) f_{\Phi \Phi} = -\frac{l^2 \omega  r_h^4}{r^4-r_h^4} \left(f_{TT}-f_{\Phi \Phi}\right).
\end{eqnarray}
We can also write $\phi \phi$ component of the metric in the off-diagonal basis can be written in terms of the metric in diagonal basis as given below:
\begin{eqnarray}
& & g_{\phi \phi}=\left(\frac{dT}{d\phi}\right)^2 G_{TT}^{\cal M}+\left(\frac{d\Phi}{d\phi}\right)^2 G_{\Phi \Phi}^{\cal M},
\end{eqnarray}
therefore $f_{\phi \phi}$ is
\begin{eqnarray}
& & f_{\phi \phi}=\left(\frac{dT}{d\phi}\right)^2 f_{TT}+\left(\frac{d\Phi}{d\phi}\right)^2 f_{\Phi \Phi} =-\left(\frac{l^2 \omega  r_h^4}{r^4-r_h^4}\right) f_{TT}+  f_{\Phi \Phi}.
\end{eqnarray}
In the small-$\omega$ limit, $\gamma=1$, therefore equation (\ref{metric-tphi-diagonal}) reduces to the following form:
\begin{eqnarray}
\label{metric-tphi-diagonal-small-l-omega}
ds^2=-\left(1-\frac{r_h^4}{r^4}\right)dT^2+ l^2 d\Phi^2.
\end{eqnarray}
Therefore ${\cal M}$-theory metric for the black hole background in the small $\omega$-limit in rotating quark-gluon plasma have the same structure as the ${\cal M}$-theory metric without rotation. Therefore higher derivative correction to the ${\cal M}$-theory metric for the rotating cylindrical black hole background will be the same as in \cite{HD-Roorkee} (in the small $\omega$-limit). Further, rotating cylindrical thermal background metric (\ref{TypeIIA-from-M-theory-Witten-prescription-T<Tc-rotating-cylindrical}) have the same structure. Therefore ${\cal O}(\beta)$ terms to the on-shell action densities for the rotating cylindrical thermal and black hole background will be similar to \cite{McTEQ} multiplied by a factor $l$, which will be canceled at the UV cut-off from both sides in equation (\ref{SBH=Sth-i}).

\end{document}